\documentclass[12pt,preprint]{aastex}

\usepackage{amsmath}

\shorttitle{A very large glitch in PSR B2334+61}
\shortauthors{Yuan et al.}

\begin{document}

\title{A very large glitch in PSR B2334+61}
\author{
J.~P.~Yuan$^{1,2}$, R.~N.~Manchester$^{3}$, N.~Wang$^{1,*}$, X. Zhou$^{1}$, 
Z. Y. Liu$^{1}$ and Z.~F.~Gao$^{1,2}$}
\affil{$^{1}$Urumqi Observatory, NAOC, 40-5 South Beijing Road,
Urumqi, Xinjiang, China, 830011} 
\affil{$^{2}$Graduate University of CAS, 19A Yuquan road, Beijing, China,
100049} 
\affil{$^{3}$ CSIRO Astronomy and Space Science, Australia Telescope National Facility, 
PO Box 76, Epping, NSW 1710, Australia} 
\affil{$^{*}$E-mail: na.wang@uao.ac.cn}

\begin{abstract}
  Seven years of pulse time-of-arrival measurements have been
  collected from observations of the young pulsar PSR B2334+61 using
  the Nanshan radio telescope of Urumqi Observatory. A phase-connected
  solution has been obtained over the whole data span, 2002 August to
  2009 August. This includes a very large glitch that occurred between
  2005 August 26 and September 8 (MJDs 53608 and 53621). The relative
  increase in rotational frequency for this glitch,
  $\Delta\nu_{g}/\nu~\sim~20.5\times10^{-6}$, is the largest ever
  seen. Although accounting for less than 1\% of the glitch, there
  were two well-defined exponential decay terms with time constants of
  21 and 147 days respectively. There was also a large long-term
  increase in the spindown rate with $\Delta\dot\nu_p/\dot\nu \sim
  0.011$ at the time of the glitch. A highly significant oscillation
  with a period of close to one year is seen in the post-glitch
  residuals. It is very unlikely that this can be accounted for by a
  pulsar position error or proper motion -- it appears to result from
  effects interior to the neutron star. Implications of these results
  for pulsar glitch models are discussed.
\end{abstract}

\section{Introduction}
Pulsars are well known for their very stable spin periods. However,
long-term timing observations of young pulsars have revealed two
categories of rotational irregularities: timing noise which is usually
characterized by continuous and erratic fluctuation in rotation
period, and glitches which are sudden increases in their rotation
rates.  Typical increases in rotational frequency during a glitch are
of the order of $\Delta\nu_{g}/\nu~\sim~10^{-9}~$ to $10^{-6}$. Large
glitches are predominately found in young pulsars with characteristic
ages between 10 kyr and 1000 kyr and are often followed by a
relaxation process in which some of the glitch decays. It
is believed that glitches and post-glitch variations reflect the
dynamics of the interior of the neutron star rather than
magnetospheric phenomena. Observing glitches and measuring their
subsequent decay processes, in addition to providing insights into the
phenomenon itself, offer one of the few probes of the neutron star
structure and thus the physics of ultra-dense matter
\citep[e.g.,][]{ll02,ab06}.

The young pulsar PSR B2334+61 (PSR J2337+6151) was discovered in the
Princeton-NRAO survey using the 92-m radio telescope at Green Bank
\citep{dtws85}; it is located near the center of, and probably
associated with, the supernova remnant G114.3+0.3 \citep{frs93}.  The
pulsar has a rotational period $P \sim 0.495$~s and a period
derivative $\dot P \sim 1.9\times10^{-13}$, implying a characteristic
age $\tau_c = P/(2\dot P) \sim 41$~kyr.  B2334+61's spin-down power,
$\dot{E}=-4\pi^{2}I\nu\dot{\nu}$, is 6.23$\times10^{34}$ erg~s$^{-1}$,
where the moment of inertia $I$ is taken to be 10$^{45}$ g
cm$^{2}$. This suggests that the pulsar may be a $\gamma$-ray emitter,
although its distance of $\sim 3$~kpc puts it at the bottom end of the
$\dot E^{1/2} d^{-2}$ metric for pulsars detectable by the {\em Fermi}
Space Telescope \citep{aaa+10c}. X-ray emission from the
pulsar was first detected in a ROSAT pointing \citep{bbt96} and was shown
to be thermal using an {\em XMM-Newton} observation \citep{mzc+06}.

In this Letter we present the timing solution of PSR B2334+61 obtained
from seven years of observation with the Nanshan 25-m telescope. The
timing analysis reveals the largest known glitch ever observed, with a
fractional frequency increase of
$\Delta\nu_{g}/\nu~\sim~20.5\times10^{-6}$, a double exponential decay
of a small part of the glitch, a large permanent increase in spindown
rate and a quasi-sinusoidal oscillation apparently triggered by the
glitch. We give a short discussion about the implications of these
results for glitch theories.

\section{OBSERVATION AND ANALYSIS}
Timing observations of PSR B2334+61 using the 25-m Nanshan radio
telescope at Urumqi Observatory commenced in 2002 July. The observing
system is described in \cite{wmz+01}. In brief, the two hands of
circular polarization at a frequency of 1540 MHz are fed through a
$2\times 128 \times 2.5$~MHz-channel filter bank and digitized at 1~ms
intervals. The observation time for this pulsar was 16 minutes.

Off-line data reduction was performed in stages using the {\sc
  psrchive} package\footnote{See http://psrchive.sourceforge.net}. The
data were dedispersed and summed to produce a total intensity
profile. Local times-of-arrival (TOAs) were determined by correlating
this profile with a standard pulse profile of high signal-to-noise
ratio. Correction of TOAs to the Solar-System barycentre and fitting
of the timing model were done using {\sc tempo2}\footnote{See
  http://www.atnf.csiro.au/research/pulsar/tempo2} with the Jet
Propulsion Laboratory DE405 ephemeris \citep{sta98b}. TOAs were
weighted by the inverse-square of their estimated uncertainty. The
basic timing model is the Taylor series
\begin{equation}\label{eq:model}
   \phi(t)=\phi_{0}+\nu(t-t_{0})+\frac{1}{2}\dot\nu(t-t_{0})^{2}+
         \frac{1}{6}\ddot\nu(t-t_{0})^{3} + \frac{1}{24}\dddot\nu(t-t_{0})^{4}
\end{equation}
where $\phi_{0}$ is the pulsar phase at the reference barycentric time
$t_{0}$.  Glitches can be modelled as combinations of changes in
$\nu$, $\dot{\nu}$ and $\ddot{\nu}$, of which parts may recover on
various timescales:
\begin{equation}\label{eq:glitch}
  \nu(t)=\nu_{0}(t) + \Delta\nu_{p}+\Delta\dot{\nu}_{p}t 
     + \frac{1}{2}\Delta\ddot{\nu}_{p}t^2 + \Delta\nu_{d}\;e^{-t/\tau_{d}}
\end{equation}
\begin{equation}\label{eq:glitch2}
 \dot{\nu}(t)=\dot{\nu}_0(t) + \Delta\dot{\nu}_p + \Delta\ddot\nu_p t +
  \Delta\dot\nu_{d}\;e^{-t/\tau_{d}}
\end{equation}
\begin{equation}\label{eq:glitch3}
 \ddot{\nu}(t)=\ddot{\nu}_0(t) + \Delta\ddot\nu_p +
 \Delta\ddot{\nu}_{d}\;e^{-t/\tau_{d}}
\end{equation}
where $\Delta\nu_p$ and $\Delta\dot\nu_p$ are permanent changes in
$\nu$ and $\dot\nu$ relative to the pre-glitch solution,
$\Delta\nu_{d}$ is the amplitude of a decaying component with a time
constant of $\tau_d$, $\Delta\dot\nu =
-\Delta\nu_d/\tau_d$ and $\Delta\ddot{\nu}_{d} =
\Delta\nu_d/\tau_d^2$. The total frequency jump at the time of the
glitch is $\Delta\nu_g = \Delta\nu_p + \Delta\nu_d$. The degree of
recovery is often described by the parameter $Q=\Delta \nu_d/\Delta
\nu_{g}$. If required, multiple decay terms with different decay time
constants can be fitted; the total frequency jump is then the sum of
the permanent and decaying frequency jumps, and similarly for
$\dot\nu$ and $\ddot\nu$.

\section{RESULTS}\label{sec:results}

The data presented here were recorded between 2002 July 31 and 2009
August 1 (MJDs 52486 and 55044). Figure~\ref{fg:dnu} gives an overview
of the results of a timing analysis, showing that a large glitch with
$\Delta\nu_{g} \approx 41\times10^{-6}$ Hz occurred between 2005
August 26 and September 7 (MJDs 53608 and 53621). The regular timing
observations of this pulsar at Nanshan allow us to track the recovery
process in detail. Most of the frequency jump $\Delta\nu_{g}$ persists
beyond the end of the data span, but the expanded $\Delta\nu$ plot
given in Figure~\ref{fg:dnu}(b) and the plot of $\dot\nu$ given in
Figure~\ref{fg:dnu}(c) show that there was an initial exponential
decay with a timescale of $\sim 150$~days and an apparently permanent
increase in slow-down rate $|\dot\nu|$. As is described in detail
below, there was an additional more rapid exponential decay with a
timescale of $\sim 21$~days following the glitch.

The rotational and positional  parameters for PSR B2334+61
from fitting of the timing model, Equation~\ref{eq:model}, to the pre-
and post-glitch data are given in Table~\ref{tb:par}. Pulsar
frequencies are in TDB units and uncertainties given in parentheses
are in the last quoted digit and are twice the values given by {\sc
  tempo2}. The position was derived from the pre-glitch data assuming
the proper motion given by \citet{hlk+04} and held fixed for the
post-glitch solution. To avoid most of the short-term post-glitch
decay, the post-glitch solution was fitted to data starting
approximately one year after the glitch. Even with this, there were
significant systematic period variations beyond the second-derivative
term, most of which were absorbed by a third-derivative term.

Figure~\ref{fg:dnu}(c) shows that there was a very
significant long-term increase in $|\dot\nu|$ at the time of the glitch;
the values of $\dot\nu$ given in Table~\ref{tb:par} reflect this
change. Furthermore, there was a significant long-term change in
$\ddot\nu$ at the time of the glitch, with the rate of decrease of
$|\dot\nu|$ increasing by more than four after the glitch. The
corresponding braking indices are $10.5\pm 0.2$ and $46.8\pm 0.3$
before and after the glitch, respectively.

\begin{figure}[h,t]
\centerline{\includegraphics[angle=-90,width=0.78\textwidth]{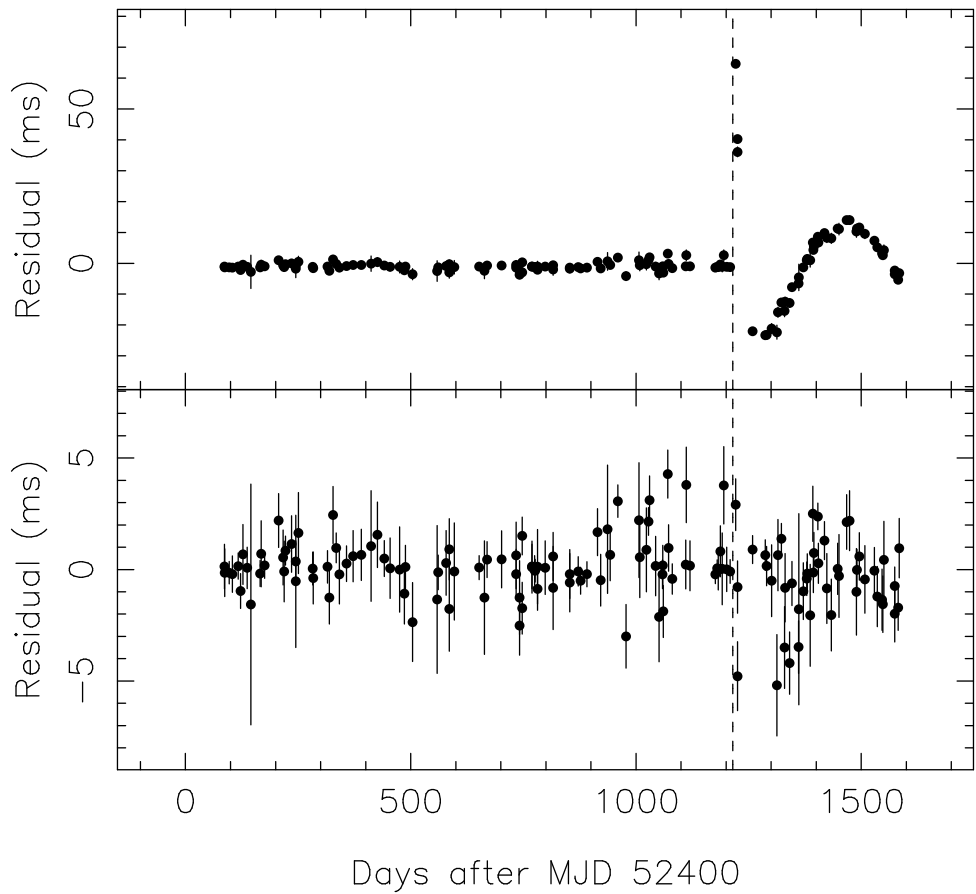}}
\caption{Timing residuals for PSR B2334+61 relative to the pre-glitch
  solution (Table~\ref{tb:par}) for data up to one year after the
  glitch. For the upper plot just one exponential decay term was
  fitted, whereas for the lower plot two exponential decay terms were
  fitted. For further details, see text. The vertical dash line indicates
the assumed epoch of glitch.}
\label{fg:decayres}
\end{figure}

Table~\ref{tb:glitch} gives glitch parameters from a fitting of a
model to the timing data. Because of the large
amplitude of the glitch, the final fit was obtained in several
stages. For the first stage, fits the data span were terminated at MJD
54000, just over a year after the glitch, to avoid contamination by
the long-term decay in $\dot\nu$. $\Delta\dot\nu_p$ was held fixed at
the value given by subtraction of the pre-glitch value of $\dot\nu$
given in Table~~\ref{tb:par} from the post-glitch value, viz,
$-7.625\times 10^{-15}$~s$^{-2}$. Then, based on Figure~\ref{fg:dnu},
a single decay with an assumed time constant of 150~d was fitted to
the TOAs, giving values of $\Delta\nu_p \sim 4.1203\times 10^{-5}$~Hz
and $\Delta\nu_{d1} \sim 1.30\times 10^{-7}$~Hz. Post-fit residuals
from this fit are shown in the upper part of
Figure~\ref{fg:decayres}. It is clear from this plot that there was an
additional initial decay with time constant $\sim 20$~d. Fitting for this with
the parameters of the longer decay held fixed gave $\Delta\nu_{d2}
\sim 1.9\times 10^{-7}$~Hz and $\tau_{d2}=18.5$~d. Residuals for this
fit shown in the lower part of Figure~\ref{fg:decayres}
demonstrate that the observed TOAs over this data span are very well
modelled by increments in $\nu$ and $\dot\nu$ together with two
exponential decays with time constants around 20~d and 150~d
respectively. 

Using these parameters as starting points, it was then possible to fit
the entire dataset from 2002 to 2009 to give the glitch parameters
listed in Table~\ref{tb:glitch}. The pulsar position, proper motion,
$\nu$, $\dot\nu$ and $\ddot\nu$ values were held fixed at the
pre-glitch values from Table~\ref{tb:par}. All other parameters were
solved for with the fit being iterated several times to ensure
convergence\footnote{Solving for the pulsar position, $\nu$, $\dot\nu$
  and $\ddot\nu$ also converged well and did not change any of the
  parameters outside their uncertainty range, demonstrating that the
  covariance between the various parameters is small.}.  \textbf{There
  was a gap of 12 days between observations around the glitch; the
  assumed glitch epoch is in the middle of this gap. Although most
  derived parameters are essentially independent of the assumed epoch,
  the amplitudes of the decaying components do change
  significantly. Quoted uncertainties in the glitch parameters include
  a contribution from the uncertainty in the glitch epoch } Residuals
from this final fit are shown in Figure~\ref{fg:allres}. Values for
$\Delta\dot\nu_p$ and $\Delta\ddot\nu_p$ are close to the differences
between $\dot\nu$ and $\ddot\nu$ for the post- and pre-glitch fits
(Table~\ref{tb:par}) as expected. They do not coincide exactly (within
the estimated errors) mainly because the exponential terms included in
the final fit absorb some of these terms.

This final fit clearly represents the glitch and the subsequent decay
extremely well. To put the scale of the residuals in
Figure~\ref{fg:allres} into context, $\Delta\nu_g$ represents a phase
gradient of one turn every 6.2 hours and the $\Delta\dot\nu_p$ and
$\Delta\ddot\nu_p$ terms correspond to a total phase contribution over
the post-glitch data span of 66.2 and 2.76 turns respectively. The
final rms residual corresponds to 0.0035 turns. Although good, the
final fit is not perfect, with a reduced $\chi^2$ of 3.32. The
$\chi^2$ is dominated by systematic deviations from the fit after the
glitch. These appear to be quasi-sinusoidal; a Lomb-Scargle spectral
analysis shows a periodicity in the post-glitch post-fit residuals of
364~d with an uncertainty of $\sim 5$~d and a significance (false-alarm
probability) of 0.0005. The fact that the period of the post-glitch
oscillations is equal to a year within the uncertainty immediately
suggests that the oscillations are due to a position error or proper
motion. However, we believe that this is not the correct
explanation. 

The position derived from the pre-glitch data (Table~\ref{tb:par}) is
fully consistent with the position extrapolated to MJD 53100 based on
the position and proper motion obtained by \citet{hlk+04} using
Jodrell Bank data spanning 16 years: 23$^{\rm h}$ 37$^{\rm m}$
05\fs779(28), +61\degr 51\arcmin 01\farcs54(17). A fit of position
only to the post-glitch data, holding all glitch and pulse frequency
parameters constant at the final-fit values, moves the position by
+0\fs059 of right ascension and $+0\farcs36$ of declination, about
four times the combined uncertainties. Compared to the extrapolated
\citet{hlk+04} position, the shifts are about twice the combined
uncertainties. Including position, proper motion, $\nu$, $\dot\nu$,
$\ddot\nu$ and all glitch terms in a fit to the whole data set
produced small changes in all glitch parameters (less than the
uncertainties given in Table~\ref{tb:glitch}) and gave a proper motion
of +111(24)~mas~yr$^{-1}$ in the right ascension direction and
+75(23)~mas~yr$^{-1}$ in declination. These values are inconsistent
with the \citet{hlk+04} values. Furthermore, the final post-fit
residuals have significant annual terms of opposite sign before and
after the glitch. These considerations strongly suggest that the
solution illustrated in Figure~\ref{fg:allres} is the best possible,
that the oscillations are confined to post-glitch epochs and that they
are not caused by a position error or proper motion of the pulsar.

There is no evidence for a change in pulse shape at the time of
the glitch, but the limits are rather poor as the typical
signal-to-noise ratio from each 16-minute observation was only about
10. 

\section{DISCUSSION}
The glitch in PSR B2334+61 is the largest in $\Delta\nu_g/\nu$ terms
(and equal largest in $\Delta\nu_g$ terms) ever observed for any radio
pulsar and probably in any pulsar (including magnetars).\footnote{See
  the Glitch Table in the ATNF Pulsar Catalogue
  (www.atnf.csiro.au/research/pulsar/psrcat).} With a characteristic
age of 41~kyr it is in the age range of pulsars known to exhibit large
glitches \citep[see, e.g.,][]{ywml10}. In fact, it lies very close in
the $P-\dot{P}$ plane to PSR J1806$-$2125 which in 1998 had a glitch
with $\Delta\nu_{g}/\nu \sim 16\times10^{-6}$ \citep{hlj+02}.  There
are other six pulsars close to PSR B2334+61 in the $P-\dot{P}$ plane:
PSRs J1551$-$5310, J1737$-$3137,
B1758$-$23, J1838$-$0453, J1841$-$0524 and B2000+32. For PSR
B1758$-$23, six glitches of moderate size ($\Delta\nu_g/\nu$ in the
range $(0.017 - 0.35)\times10^{-6}$) have been detected. For PSRs
J1737$-$3137 and J1841$-$0524 a single glitch of relative size $\sim
1.3\times10^{-6}$ and $1.0\times10^{-6}$ respectively have been
reported \citep{wjm+10}. Other pulsars have not been reported to
glitch so far.  This is the first glitch reported for PSR B2334+61;
since the pulsar was observed for 16 years from 1987 by
\citet{hlk+04}, the time between glitches is at least 22 years.

In contrast to the case for PSR J1806$-$2125, for PSR B2334+61
observations were obtained close to the date of the glitch allowing a
detailed examination of the post-glitch behaviour. 
PSR B2334+61 continues the trend for minimal post-glitch
recovery of $\Delta\nu_{g}$, i.e., small $Q$, for large glitches in
middle-aged pulsars \citep{lsg00,ywml10}. Despite this, the good sampling of
the post-glitch behaviour shows clear evidence for two distinct
exponential recoveries and, together with permanent jumps in $\dot\nu$
and $\ddot\nu$, these provide a remarkably accurate description of the
post-glitch behaviour. Previously, only the Vela pulsar has shown
clear evidence for multiple exponential decays associated with a given
glitch \citep{dml02}. For Vela, the longest decay timescale
observed by \citet{dml02} is about 19~d, comparable to the shorter
decay timescale observed for PSR B2334+61. There is evidence
for longer exponential decays in the Vela pulsar glitches which
ultimately become more linear \citep{lpgc96,lsg00,wmp+00}. 

The standard model for large pulsar glitches envisages a sudden
transfer of angular momentum from a superfluid component to the rest
of the star -- the vortex unpinning model \citep{ai75,als84}. The
theoretical understanding of glitches includes the trigger mechanism
for the glitch, the strength of the vortex pinning and the post-glitch
evolution. \citet{ll02} investigated a model in which vortex unpinning
results from a sudden heating of the star, possibly due to a starquake
or crustal movement, whereas \cite{ga09} suggested that an r-mode
instability may trigger a global unpinning of vortices leading to a
glitch. \citet{accp93} and earlier papers by these authors attribute
the long-term post-glitch relaxation to vortex creep, with different
regions of the star having different properties to account for the
sometimes complicated post-glitch behaviour. \citet{rzc98} suggested
that crustal motions could account for essentially all glitch
properties; see also \citet{rud09}.

The fractional jump in $\dot\nu$ at the time of the glitch was very
large, about 15\% (Table~\ref{tb:glitch}), much larger than the value
of 1.7\% found by \citet{lsg00} to represent most pulsars.  For PSR
B2334+61, $\Delta\dot\nu_g/\dot\nu$ is dominated by the fastest decay
term, with the relative contributions for the 21-d decay, the 147-d
decay and the permanent jump being 85\%, 8\% and 7\%, respectively. In
most glitch theories, $\Delta\dot\nu_g/\dot\nu$ represents the moment
of inertia fraction which is in a superfluid state not tightly coupled
to the neutron-star crust \citep[see, e.g.,][]{acp89,accp93}. The
large value for PSR B2334+61 suggests that a considerable fraction of
the star is weakly coupled superfluid, with most of it returning to an
equilibrium rotational state within about 20 days. About
1\% of the rotational inertia appears to be more loosely coupled,
leading to the 147-d decay, and a further 1\% barely coupled at all.

\begin{figure}[h,t]
\centerline{\includegraphics[angle=-90,width=0.65\textwidth]{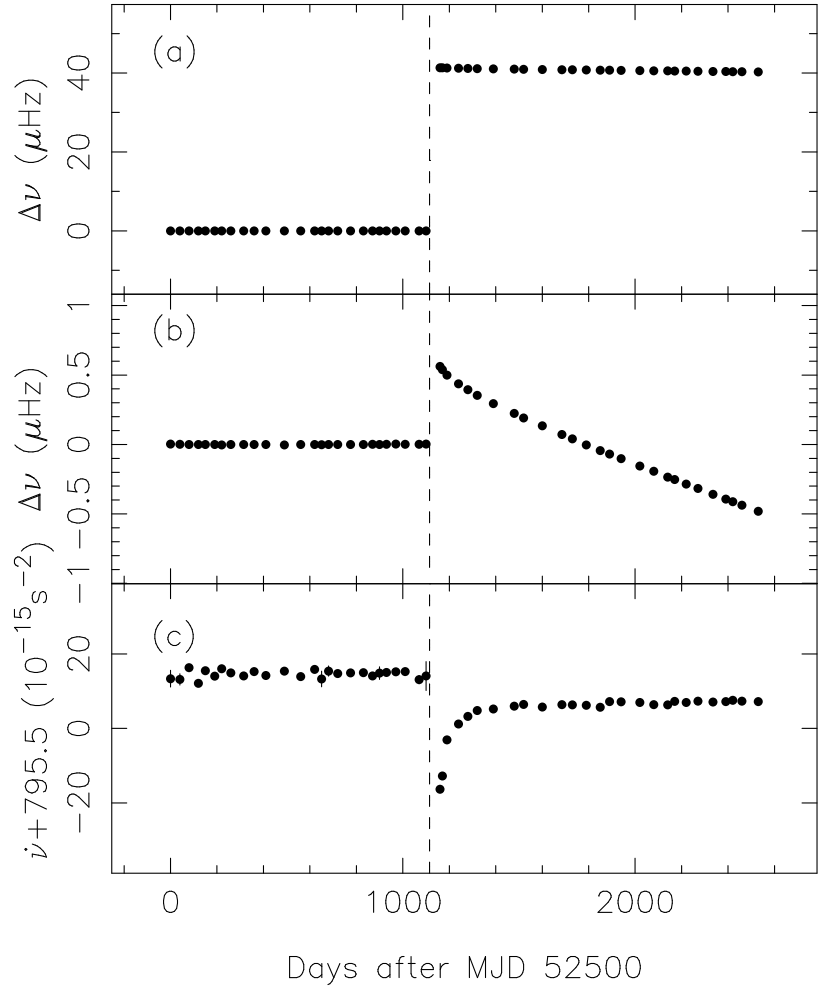}}
\caption{Glitch of PSR B2334+61: (a) variations of rotational
  frequency $\Delta\nu$ relative to the pre-glitch solution, (b) an
  expanded plot of $\Delta\nu$ where the mean post-glitch value has
  been subtracted from the post-glitch data, and (c) variations of the
  frequency first derivative $\dot{\nu}$. The vertical dashed line
  marks the glitch epoch. The plotted points represent fits of $\nu$
  and $\dot\nu$ to 5 -- 10 adjacent TOAs, with $\Delta\nu$ being the
  difference between the fitted value of $\nu$ and the value from the
  pre-glitch solution (extrapolated after the glitch). }
\label{fg:dnu}
\end{figure}

The very significant quasi-permanent change in $\dot\nu$ associated
with the PSR B2334+61 glitch (Figure~\ref{fg:dnu}) is unusual. Such
clear quasi-permanent changes in $\dot\nu$ have previously only been
observed in the Crab pulsar \citep{lps93}. Although it is fairly
common for only part of $\Delta\dot\nu_g$ to recover exponentially
after a large glitch, in most cases there is a continuing
approximately linear recovery of $\dot\nu$. As discussed above, this
was seen in the Vela pulsar period, and other examples have been
presented and discussed by \citet{lsg00} and \citet{ywml10}. For PSR
B2334+61, the fractional change $\Delta\dot\nu_p/\dot\nu$ is about
1.1\%, about 25 times larger than the value of $\sim 4\times 10^{-4}$
observed for the Crab pulsar. These sudden changes in $\dot\nu$
implies equally sudden changes in the effective braking torque at the
time of the glitch since a change in the stellar moment of inertia of
this magnitude is ruled out by the fact that $\Delta\nu_p/\nu$ is
several orders of magnitude smaller. The observed change in $\dot\nu$
for PSR B2334+61 seems too large to be accounted for by exterior
magnetic field changes as suggested for the Crab pulsar
\citep{rzc98,rud09}. Furthermore, there was no major change in the
pulse shape at the time of the glitch, suggesting that magnetospheric
changes are not responsible. Therefore it is most likely that changes
interior to the neutron star are responsible. The effective braking
torque is the difference between magnetospheric torques and internal
torques believed to be due to vortex creep
\citep[e.g.,][]{aaps84a}. This model predicts a linear increase in
$\dot\nu$ over long timescales as is seen in the Vela pulsar
\citep{accp93}. For PSR B2334+61 the observed value of $\ddot\nu$
is two orders of magnitude smaller than for Vela, which is in
accordance with the predictions of the vortex-creep model \citep{ab06}.

The observed long-term step in $\ddot\nu$ (Table~\ref{tb:glitch})
could be interpreted as indicating a third exponential decay with a
very long timescale. Large positive values of $\ddot\nu$ are commonly
attributed to the long-term effect of (sometimes unseen) glitches
\citep[e.g.,][]{hlk+04,ab06}. For an exponential decay,
$\Delta\ddot\nu_{d3} \sim \ddot\nu_p =
\Delta\nu_{d3}/\tau_{d3}^2$. Since we do not know $Q_{d3} =
\Delta\nu_{d3}/\Delta\nu_g$, we can just set a limit on $\tau_{d3} <
70$~years corresponding to $Q_{d3} = 1$. In principle, the observed
post-glitch value of $\dddot\nu$ (Table~\ref{tb:par}) could be
interpreted as $\Delta\dddot\nu_{d3}$ and used to solve for $Q_{d3}$
and $\tau_{d3}$, but it is inconsistent with this, implying a decay
timescale $-\Delta\ddot\nu_{d3}/\Delta\dddot\nu_{d3}$ of less than one
year. Evidently the shorter-term decays are not exactly exponential or
other noise processes contribute to the observed value of $\dddot\nu$.

\begin{figure}[h,t]
\centerline{\includegraphics[angle=-90,width=0.78\textwidth]{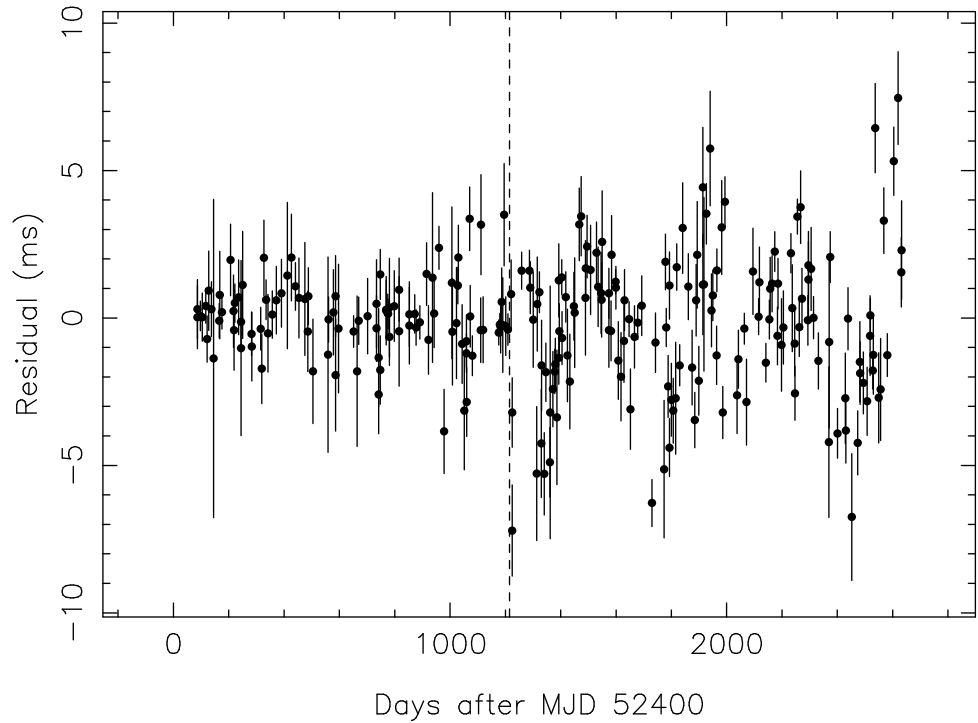}}
\caption{Timing residuals for the entire seven-year dataspan after
  fitting for two exponential recoveries and a permanent changes in
  $\dot\nu$ and $\ddot\nu$. The pulsar position and frequency
  parameters were held fixed at values obtained from the pre-glitch
  solution. }
\label{fg:allres}
\end{figure}

The oscillations in the post-glitch timing residuals are a remarkable
feature of the timing behaviour of PSR B2334+61
(Figure~\ref{fg:allres}). These oscillations have a period very close
to one year, but in Section~\ref{sec:results} we showed that an
interpretation in terms of position error or proper motion is very
improbable. Quasi-periodic oscillations in timing residuals have been
seen in the Crab pulsar \citep{lps93} and many other pulsars
\citep{hlk10}. In none of these cases though are the oscillations
clearly associated with a glitch although glitches have been mentioned
as an excitation mechanism \citep[e.g.,][]{zxw+04,tim07}. The
observed quasi-periodic oscillations have often been interpreted as
due to Tkachenko oscillations in the superfluid vortex array
\citep{rud70,pop08}. Under certain reasonable assumptions, these
oscillations have a period given by
\begin{equation}
P_T \sim 1.77 R_6 P^{1/2} {\rm yr}
\end{equation}
where $R_6$ is the neutron-star radius in units of $10^6$~cm and $P$
is the pulsar period in seconds. For PSR B2334+61, this gives a period
of $1.24 R_6$~yr, remarkably close to the observed period of 1
yr. This suggests that Tkachenko
oscillations are a viable mechanism for the observed
oscillations. Further investigation is needed to understand the
triggering mechanism and whether or not the observed oscillation
amplitude can be accounted for.

\begin{acknowledgements}
We thank George Hobbs for helpful discussions. This work is supported by 
National Basic Research Program of 
China $-$ 973 Program 2009CB824800, NSFC projects 10673021, 10778631 and 10903019,  
Knowledge Innovation Program of The Chinese Academy Sciences  KJCX2-YW-T09, West Light Foundation of  CAS  (No. XBBS200920) and Xinjiang Natural 
Science Foundation (No. 2009211B35).
\end{acknowledgements}

%\bibliographystyle{apj}
%\bibliography{journals,modrefs,psrrefs,crossrefs}

\clearpage

\begin{deluxetable}{lcc}
\tablecaption{Pre- and post-glitch timing parameters for PSR B2334+61\label{tb:par}}
\tablehead{\colhead{Parameter} & \colhead{Pre-glitch} & \colhead{Post-glitch}}
\startdata
Right ascension (J2000)             &  \multicolumn{2}{c}{23$^{\rm h}$ 37$^{\rm m}$ 05\fs762(10)} \\
Declination (J2000)                 &  \multicolumn{2}{c}{+61\degr 51\arcmin 01\farcs53(7)} \\
Proper motion in RA dirn (mas yr$^{-1}$) & \multicolumn{2}{c}{$-1$\tablenotemark{a}} \\
Proper motion in Dec. (mas yr$^{-1}$) & \multicolumn{2}{c}{$-15$\tablenotemark{a}} \\
Epoch of position  (MJD)            &  \multicolumn{2}{c}{53100}\\
& &\\
Frequency, $\nu$ (Hz)                &  2.01874888482(2)   & 2.01869363589(2)   \\
Frequency derivative, $\dot\nu$ (10$^{-15}$~s$^{-2}$)      &  $-$780.7074(5) & $-$788.332(3)  \\
Frequency 2nd derivative, $\ddot\nu$ (10$^{-24}$~s$^{-3}$)   &  3.17(6)  & 14.46(9)  \\
Frequency 3rd derivative, $\dddot\nu$ (10$^{-30}$~s$^{-4}$)   & \nodata  & $-$0.37(2)  \\
Frequency epoch (MJD)               &       53100.0       & 54521.0 \\
%Epoch (MJD)               &       53100.0       & 54521.0 \\
Data span (MJD)                     & 52486--53609 & 53998--55045 \\
DM (pc cm$^{-3}$)                   &         58.410   &  58.410 \\
No. of TOAs                                &                98  & 100\\
RMS timing residual (ms)            &         0.91   & 2.07 \\
Reduced $\chi^2$                   & 1.04  & 4.88 \\
\enddata
\tablenotetext{a}{From \citet{hlk+04}}
\end{deluxetable}

\begin{deluxetable}{lc}
\tablecaption{The glitch parameters\label{tb:glitch}}
\tablehead{\colhead{Parameter} & \colhead{Value}}
\startdata
Glitch epoch (MJD)                                & 53615(6) \\
$\Delta\nu_g/\nu$ (10$^{-6}$)                & 20.5794(12) \\
$\Delta\dot\nu_g/\dot{\nu}$      & 0.156(4)  \\
$\Delta\dot\nu_p$  ($10^{-15}$~s$^{-2}$)     & $-$8.684(17) \\
$\Delta\ddot\nu_p$ ($10^{-24}$~s$^{-3}$)      & 8.4(2)  \\
$\Delta\nu_{d1}$   ($\mu$Hz)                & 0.19(3)   \\
$\tau_{d1}$  (d)                           & 21.4(5)  \\
$\Delta\nu_{d2}$   ($\mu$Hz)                  & 0.119(4)       \\
$\tau_{d2}$   (d)                        & 147(2)   \\
$Q=(\Delta\nu_{d1}+\Delta\nu_{d2})/\Delta\nu_g$    & 0.00751(5) \\
Data span (MJD)                           & 52486--55045 \\
Rms timing residual (ms)                            & 1.72 \\
Reduced $\chi^2$                          & 3.32 \\
\enddata
\end{deluxetable}

\label{lastpage}
\end{document}